\documentclass[12pt,3p]{nj_style}

\usepackage{amsmath}
\usepackage{amssymb}
\usepackage{amsthm}
\usepackage{enumerate}
\usepackage[font=small,labelfont=bf,margin=8pt]{caption}
\usepackage[usenames,dvipsnames]{color}
\usepackage{graphicx}
\usepackage[colorlinks=true]{hyperref}
\usepackage{tikz}
\tikzstyle{vertex}=[circle, draw, inner sep=0pt, minimum size=4pt]

\newtheorem{theorem}{\bf Theorem}[section]

\newtheorem{proposition}{Proposition}
\newtheorem{observation}{Observation}

\newcommand{\ket}[1]{|#1\rangle}

\newcommand{\ip}[2]{\langle #1|#2\rangle}
\newcommand{\op}[2]{|#1\rangle \langle #2|}

\newcommand{\tr}{{\rm tr}}

\usepackage{amsthm}
\usepackage{dcolumn}
\usepackage{bm}
\usepackage{graphicx}
\usepackage{color}

\usepackage{listings}

\makeatletter
\renewcommand*\env@matrix[1][c]{\hskip -\arraycolsep
  \let\@ifnextchar\new@ifnextchar
  \array{*\c@MaxMatrixCols #1}}
\makeatother

\begin{document}

\begin{frontmatter}


\title{Characterization of multipartite entanglement \\ in terms of local transformations}
\author[UTS]{Youming Qiao}
\ead{Youming.Qiao@uts.edu.au}

\author[ICT]{Xiaoming Sun}
\ead{sunxiaoming@ict.ac.cn}

\author[UTS,IQC,UG]{Nengkun Yu}
\ead{nengkunyu@gmail.com}

\address[UTS]{Centre for Quantum Software and Information,\\
   Faculty of Engineering and Information Technology, \\ University of
   Technology Sydney, NSW 2007, Australia}
\address[ICT]{Institute of Computing Technology, Chinese Academy of Sciences}
\address[IQC]{Institute for Quantum Computing, University of Waterloo, Waterloo, Ontario N2L~3G1, Canada}
\address[UG]{Department of Mathematics \& Statistics, University of Guelph, Guelph, Ontario N1G~2W1, Canada}

\begin{abstract}

The degree of the generators of invariant polynomial rings of is a long standing open problem since the very initial study of the invariant theory in the 19th century. Motivated by its significant role in characterizing multipartite entanglement, we study the invariant polynomial rings of local unitary group---the tensor product of unitary group, and local general linear group---the tensor product of general linear group. For these two groups, we prove polynomial upper bounds on the degree of the generators of invariant polynomial rings. On the other hand, systematic methods are provided to to construct all homogenous polynomials
that are invariant under these two groups for any fixed degree. Thus, our results can be regarded as a complete characterization of the invariant polynomial rings. As an interesting application, we show that multipartite entanglement is additive in the sense that two multipartite states are local unitary equivalent if and only if $r$-copies of them are LU equivalent for some $r$.
\end{abstract}


\end{frontmatter}

\section{Introduction}

Multipartite entanglement is considered as an essential asset to quantum information
processing and computational tasks. The intriguing properties and potential applications of entanglement spark many literature dedicated to quantify it as a resource. Even though great efforts and considerable progress have been made
\cite{BC01,KEM99,DVC00,EB01,HEB04,VDMV02,GW11,GW10,YDY10,YCGD10,YGD13,YU13,MV04}, no complete theory can be obtained.

The first approach for study multipartite entanglement is to study the local unitary(LU) equivalence of multipartite states. The importance of this approach is due to the fact that multipartite entanglement is characterized by the equivalence
relation under LU. Bennett $et.al.$ proved the important fact that two quantum states are interconvertible by unlimited
two-way classical
communication (LOCC) if and only if they are interconvertible by
LU \cite{BPR+00}. The celebrated Schmidt decomposition provides the canonical form for bipartite pure states under LU, which enables us to understand bipartite entanglement completely. For multipartite system, no such decomposition is possible. Hence, understanding multipartite entanglement is much more challenging.
Lots of efforts have been made to study the LU equivalent relation, see
\cite{SZA12,HW09,HWK09,VRA11a,VRA11b,AFPY03,WWWF11,ZZF+12,OD09,KRA10a,KRA10b} as a very incomplete list.

In principle, this LU equivalent relation can be characterized by the ring of invariants polynomials under local unitary(LUIPs) \cite{RAI00,MW02,GRB98}.
From this point of view, multipartite entanglement is characterized by the ring of LUIPs. A complete description of such ring has been obtained only for bipartite system and $2\times 2 \times n$ system \cite{MAK02,SUD01,WAL05}. Beyond that, and despite the extensive
literature, very little is known. One related topic in the study of quantum information science is the stochastic local operations and classical communication(SLOCC). In 2013, Gour and Wallach
constructed the whole set of SL-invariant polynomials (SLIPs) for pure states \cite{GW13}.

For the ring of LUIPs, it is already
known that the ring of LUIPs is
finitely generated.
In order to understand the the ring of LUIPs ring, two central problems have to be
addressed. The first is to construct the ring by presenting some finite generating set of LUIPs. The second problem
naturally arises after the first: bound the degree of generating set, more precisely, present an explicit upper bound on the degree such that the ring of LUIPs
can be generated by the LUIPs with degree lower than that bound.

In this paper, we give a characterization of multipartite entanglement by
conquering the above two problems. We first provide an algorithm to construct
all LUIPs for fixed degree.
Secondly, we demonstrate an explicit polynomial upper bound on the degree of generators by employing
modern techniques and concepts of invariant theory. As far as we know, no explicit degree bound for the
ring of LUIPs was computed in the literature. As an interesting application, we are able
to show that multipartite entanglement is additive in the sense that two
multipartite states are LU equivalent if and only if $r$-copies of these two
states are LU equivalent for some $r$.

Our main idea to construct LUIPs here is remarkably simple and feasible in all dimensions: For each party, we construct the corresponding ring of homogenous
polynomials that are invariant under the local unitary group applied on that
party, then the ring of LUIPs is the intersection of these rings, which
can be obtained by computing the intersection of finite dimensional subspaces. To prove a polynomial upper bound on the degree of generators, new techniques on matrix semi-invariants \cite{DM15,IQS15} are employed.

This method can also be used to study the SLOCC equivalence of multipartite states. For pure
states, we provide an alternative algorithm to compute SLIPs rather than
\cite{GW13}. For mixed states, we focus on the so called one term SLOCC equivalence relation.

\section{Preliminaries and Notations}
\subsection{Notations}
During this paper, we consider the $n$-partite Hilbert space
\begin{equation*}
\mathcal{H}=\mathcal{H}_1\otimes\mathcal{H}_2\otimes\cdots\otimes\mathcal{H}_n,
\end{equation*}
where Hilbert space $\mathcal{H}_i$ has dimension $d_i$. Then the dimension of $\mathcal{H}$ is $\Pi_{i=1}^n d_i$.

Let $\mathbb{U}(s)$ be the group of $s\times s$ unitary matrices. Then local
unitary(LU) group is defined as
\begin{equation*}
\mathbb{LU}\equiv\mathbb{U}(d_1)\otimes\mathbb{U}(d_2)\otimes\cdots\otimes\mathbb{U}(d_n).
\end{equation*}
For $\ket{\Psi}\in \mathcal{H}$, the orbit
$\mathbb{LU}\ket{\Psi}:=\{g\ket{\Psi}:g\in \mathbb{LU}\}$ consists of
quantum states in the LU equivalent class of $\ket{\Psi}$. The orbit characterizes the multipartite entanglement in the sense that the entanglement of $\ket{\Psi}$ is the same as that of any state in $\mathbb{LU}\ket{\Psi}$. Therefore, the goal of characterizing multipartite entanglement can be accomplished by separating different LU orbits, in other words, determining the LU equivalence relation.

This problem can be formalized as: For two given states $\ket{\Psi},\ket{\Phi}\in\mathcal{H}$, determine whether there exists $g\in \mathbb{LU}$ such that $$\ket{\Psi}=g\ket{\Phi}.$$
To demonstrate the importance of LU equivalence of pure states, we would like
to point out the following fact that the problem of the LU equivalence of mixed
states can be reduced to the same problem of pure states, $i.e.$, two
$n$-partite mixed states are LU equivalent if and only if their purifications, two $n+1$-partite pure states, are LU equivalent \cite{AFPY03}, where two mixed states $\rho,\sigma$ in Hilbert space $\mathcal{H}$ are called LU equivalent if there exists $(U_1\otimes U_2\otimes\cdots\otimes U_n)\in \mathbb{LU}$ such that $\rho=(U_1\otimes U_2\otimes\cdots\otimes U_n)\sigma(U_1\otimes U_2\otimes\cdots\otimes U_n)^{\dag}$.

The LU equivalence of mixed states can be used to study the equivalence between quantum channels, where two quantum channels $\mathcal{E}$ and $\mathcal{F}$ are said to be equivalent if there are unitary channels $\mathcal{U}$ $\mathcal{V}$ such that
\begin{equation*}
\mathcal{F}=\mathcal{V}\circ \mathcal{E}\circ\mathcal{U}.
\end{equation*}
Here, unitary channels $\mathcal{U}$ and $\mathcal{V}$ can be regarded as encoding channel and decoding channel, respectively. It is direct to verify that
$\mathcal{E}$ and $\mathcal{F}$ have the same ability on transmit information,
quantum (classical, private) capacity.
One can observe that $\mathcal{E},\mathcal{F}:L(\mathcal{H_A}):\mapsto
L(\mathcal{H_B})$ are equivalent if and only if their Choi-matrices are LU
equivalent, where the Choi-matrix of
$\mathcal{E}$ is defined as the bipartite mixed
state (non-normalized) $\rho_{AA'}=(\mathcal{I}_{A'}\otimes
\mathcal{E})(\op{\varphi}{\varphi})$ with $\ket{\varphi}=\sum_{j=1}^d \ket{i}\ket{i}$, and
the noiseless channel $\mathcal{I}_{A'}$ on quantum system $\mathcal{H}_{A'}$
which has the same dimension $d$ as system $\mathcal{H}_{A}$.

Another widely studied equivalence relation is the SLOCC equivalence. Two pure states $\ket{\Psi},\ket{\Phi}\in\mathcal{H}$ are called SLOCC equivalent if there is some $g\in\mathbb{G}$ and $\lambda\in\mathbb{C}$ such that $$g\ket{\Psi}=\lambda \ket{\Phi}$$ with $\mathbb{G}=\mathbb{SL}(d_1,
\mathbb{C})\otimes \dots \otimes \mathbb{SL}(d_n, \mathbb{C})$ and $\mathbb{SL}(d_i, \mathbb{C})$ standing for the set of $d_i\times d_i$ invertible matrices with determinant 1.

This problem becomes much more complicated for mixed states: Two mixed states $\rho,\sigma$ in Hilbert space $\mathcal{H}$ are called SLOCC equivalent if there exists $h_i=(A_{i,1}\otimes A_{i,2}\otimes\cdots\otimes A_{i,n})$ and $m_j=(B_{j,1}\otimes B_{j,2}\otimes\cdots\otimes B_{j,n})$ such that
\begin{align*}
\rho=\sum_{i}h_i\sigma h_i^{\dag},\\
\sigma=\sum_{j} m_j\rho m_j^{\dag}
\end{align*}
with $A_{i,k}$ and $B_{j,k}$ being $d_j\times d_j$ matrices for all $i,j,k$.

To determine the SLOCC equivalence between mixed states, even between pure state and mixed state, becomes very difficult. To see this, we notice that $\ket{0}\ket{0}\cdots\ket{0}$ is SLOCC equivalent to any separable states. That is, to see whether a given state is SLOCC equivalent to $\ket{0}\ket{0}\cdots\ket{0}$, we need to test whether it is separable, that problem has been widely studied and it is known to be NP-Hard \cite{GUR03}. To simplify the problem of SLOCC equivalence, we focus on the so called on term SLOCC equivalence. Two mixed states $\rho,\sigma$ are called one term SLOCC equivalent if there exists $g\in\mathbb{G}$ and $\lambda\in\mathbb{C}$ such that
\begin{align*}
\rho=\lambda~g\sigma g^{\dag}.
\end{align*}
\subsection{Invariant Polynomials}
In this subsection, we demonstrate some basic notions of invariant polynomials under LU and under SLOCC, respectively.

We use the concept of LUIPs, namely the polynomials invariant
under local unitary transformations, to study the LU equivalence. Formally, a
function $f:\mathcal{H}\mapsto \mathbb{C}$ is an LUIP, if $f(\ket{\Psi})$ is
the homogenous polynomial on entries of $\op{\Psi}{\Psi}$($\Psi$ in short), and
\begin{equation*}
f(g\ket{\Psi})=f(\ket{\Psi}), ~~\forall g\in \mathbb{LU}~~\mathrm{and}~~\forall\ket{\Psi}\in \mathcal{H}.
\end{equation*}
Notice that any polynomial can be written as a linear combination of homogenous polynomials, and the invariance follows naturally.

Let $\mathbb{C}[\mathcal{H}]^{\mathbb{LU}}$ denote the set of the LUIPs. It is direct to see that $\mathbb{C}[\mathcal{H}]^{\mathbb{LU}}$ is a ring. In other words, it is an abelian group under addition, a monoid under multiplication. Moreover, polynomial multiplication is distributive with respect to addition.

It is well known that the entanglement of bipartite state $\ket{\Psi}_{AB}$ is completely determined by its vector of Schmidt
coefficients, says $(\lambda_1,\cdots,\lambda_d)$($\lambda_1\geq\cdots\geq\lambda_d$), or equivalently determined by $\sum_{j=1}^{d}\lambda_j^{k}=\tr(\Psi_A)^k$ for $k\in \mathbb{N}$. By noticing that $\tr(\Psi_A)^k$ is value of LUIP for $\ket{\Psi}_{AB}$, we know that two bipartite states are LU equivalent if and only if LUIPs have the same value for them.

In general, multipartite quantum states $\ket{\Psi}$ and $\ket{\Phi}$ are LU equivalent if and only if $f(\ket{\Psi})=f(\ket{\Phi})$ holds for every LUIP $f$.

SL-invariant polynomials (SLIPs) can be used to study the SLOCC equivalence between quantum states.

To see the power of our method on constructing LUIPs, we apply it on studying the SLOCC equivalence where
SLIP is a polynomial $f:\mathcal{H}\mapsto \mathbb{C}$ such that
\begin{equation*}
f(g\ket{\Psi})=f(\ket{\Psi}), ~~\forall g\in \mathbb{G}~~\mathrm{and}~~\forall\ket{\Psi}\in \mathcal{H}.
\end{equation*}

Very recently, Gour and Wallach present an algorithm for constructing the SLIPs for fixed degree using Schur-Weyl duality.

\section{Main results}\label{sec:proof}

In this section, we first provide a new view of LUIPs
which leads to an algorithm  to compute the ring of LUIPs. We also demonstrate
an explicit upper bound such that any LUIP can be
generated, using addition, subtraction and product, by LUIPs with degrees no more
than the bound. Based on this characterization, we are able to show that two
multipartite states are LU equivalent if and only if $r$-copies of these two
states are LU equivalent for some $r$.

Let $I_j$ be the identity operator of system $\mathcal{H}_j$, and $\mathbb{I}_i=\{I_j\}$, we can define group $\mathbb{U}_i$ as follows,
\begin{equation*}
\mathbb{U}_i=\mathbb{I}_1\otimes \cdots\otimes \mathbb{I}_{i-1}\otimes\mathbb{U}(d_i)\otimes \mathbb{I}_{i+1}\cdots\otimes \mathbb{I}_n.
\end{equation*}
A useful observation is
\begin{equation*}
\mathbb{U}_i\subset
\mathbb{LU},\mathrm{and}~~\mathbb{LU}=\mathbb{U}_1\mathbb{U}_2\cdots \mathbb{U}_n.
\end{equation*}
The advantage of this observation on studying the polynomial invariants is based on the following relation between the polynomial invariants of $\mathbb{LU}$, says $P$, and those polynomial invariants of $\mathbb{U}_i$s, says $P_i$s,
\begin{equation}\label{basic}
P=\bigcap_{i=1}^n P_i.
\end{equation}
First, we observe that $P\subset P_i$ by noticing $\mathbb{U}_i\subset \mathbb{LU}$. Thus, $P\subset \bigcap_{i=1}^n P_i$.

On the other hand, one can verify that for any $p\in \bigcap_{i=1}^n P_i$, $g=g_1g_2\cdots g_n\in \mathbb{LU}$ with $g_i\in \mathbb{U}_i$ and $\ket{\varphi}\in\mathcal{H}$, we have $p\in P$ by observing
\begin{eqnarray*}
&&p(g\ket{\varphi})\\
&=&p(g_1\cdots g_n\ket{\varphi})\\
&=&p(g_2\cdots g_n\ket{\varphi})\\
&=&\cdots\\
&=&p(\ket{\varphi}).
\end{eqnarray*}
Therefore, $P\supset \bigcap_{i=1}^n P_i\Rightarrow P=\bigcap_{i=1}^n P_i$.

We only need to compute $P_i$ for fixed $i$, the ring of invariant polynomials under unitary group $\mathbb{U}_i$. To study the action of $\mathbb{U}_i$ on $\mathcal{H}$, one may regard the whole space $\mathcal{H}$ as a bipartite space: system $\mathcal{H}_i$ and the rest. Now, the problem becomes to compute the invariants of one party unitary for bipartite pure state. Formally,
suppose $\ket{\varphi}=\sum x_{j_1\cdots j_n}\ket{j_1\cdots j_n}$ with variables $x_{j_1\cdots j_n}\in\mathbb{C}$.
According to Uhimann's theorem \cite{Uhl76}, the set of the entries of $\varphi_{\bar{i}}=Tr_i \op{\varphi}{\varphi}$, those quadratic polynomials, form a generating set of $P_i$.

For any degree $l$, the relation (1) enables us to compute the whole set of degree
$l$ homogenous elements of $P$. One can verify that any LUIP is of even degree.

In order to accomplish the characterization of LUIPs, we need the following
theorem which explicitly provides an upper bound on the degree to generate the
ring of LUIPs.

\begin{theorem}
The set of all LUIPs is generated by the LUIPs with degree no more than
$N(d_1,d_2,\cdots,d_n)=\frac{3}{8}(\prod_i
d_i^2)\cdot \max(d_i)^{2n}\cdot (\sum_i d_i)^{2(\sum_i d_i^2-n)}$.
\end{theorem}
\textit{Remark:---}Although it is known that the ring of LUIPs is finitely
generated, as far as we know, no explicit bound was reported in the study of
quantum information
theory. In modern invariant theory, such an explicit bound for the degree of
generating set is known for linearly reductive algebraic group acting rationally
on linear spaces \cite{DER01}. Here, the LU group is not a linearly reductive
algebraic group. In order to apply this theory on our problem, we resort to the
concept ``complexification" from \cite{Naimark}. For readability, we postpone the
detailed proof of
this conclusion to section 4. An upper bound for SLIPs is also included, whose
detailed definition is given in the later of this paper. Note that the bound for
SLIPs was not computed in the literature before this work.

Thus, two quantum states $\ket{\Psi},\ket{\Phi}\in\mathcal{H}$ are LU equivalent if and only if $f_i(\ket{\Psi})=f_i(\ket{\Phi})$ holds for a basis LUIPs $f_i$ with degree less than $N(d_1,d_2,\cdots,d_n)$.

This characterization of LUIPs can be regarded as a demonstration of the
decidability of LU equivalence. Although this fact can also be observed according
to Tarski-Seidenberg's famous result, our result is still valuable since it
contains physical background and can bring new insight of the entanglement while
Tarski-Seidenberg's result does not provide. As an illustration, the following
proof of the additivity of entanglement crucially depends on the structure of
LUIPs provided above.

Consider the LU equivalence of the $r$-copy set of states $\mathcal{H}^r:=\{\ket{\varphi}^{\otimes r}:\ket{\varphi}\in \mathcal{H}=\bigotimes_{i=1}^n\mathcal{H}_i\}$, where $\ket{\varphi}^{\otimes r}$ is regarded as $n-$partite state of system $\bigotimes_{i=1}^n\mathcal{H}_i^{\otimes r}$. Thus, $\ket{\Psi}^{\otimes r},\ket{\Phi}^{\otimes r}\in \mathcal{H}^r$ are called LU equivalent if there exists local unitary $\bigotimes_{i=1}^n U_i$ with $U_i$ being unitaries of system $\mathcal{H}_i^{\otimes r}$ such that $\ket{\Psi}^{\otimes r}=\bigotimes_{i=1}^n U_i\ket{\Phi}^{\otimes r}$.

One can observe that if $\ket{\Psi},\ket{\Phi}$ are LU equivalence then
$\ket{\Psi}^{\otimes r},\ket{\Phi}^{\otimes r}$ are LU equivalence in the above
sense. Here, we are interested in the converse direction.

For the bipartite case, one can conclude that the converse is also true by using
the following argument: Without loss of generality, assume
$\ket{\Psi}_{AB}=\sum_{j=1}^{d}\sqrt{\lambda_j} \ket{jj}$ and
$\ket{\Phi}_{AB}=\sum_{j=1}^{d}\sqrt{\delta_j} \ket{jj}$. Since
$\ket{\Psi}_{AB}^{\otimes r},\ket{\Phi}_{AB}^{\otimes r}$ are LU-equivalent, we
know that $\Psi_{A}^{\otimes r}$ and $\Phi_{A}^{\otimes r}$ share eigenvalues with
$\Psi_{A}$ and $\Psi_{A}$ being the reduced density matrices of $\ket{\Psi}_{AB}$
and $\ket{\Phi}_{AB}$ respectively.
Thus, $\sum_{j=1}^{d}\lambda_j^{rk}=\sum_{j=1}^{d}\delta_j^{rk}$ for all $k$, and
one can conclude that the Schmidt coefficients $\lambda_i$s are identical with
$\delta_i$s. That is, $\ket{\Psi},\ket{\Phi}$ are LU-equivalent.

Interestingly, the converse is also true for general multipartite system. To prove
such a claim, we need to use a new tool rather than Schmidt coefficients--the LUIPs.
\begin{theorem}
If $\ket{\Psi}^{\otimes r},\ket{\Phi}^{\otimes r}$ are LU equivalent for some
$r\in\mathbb{N}$, then $\ket{\Psi},\ket{\Phi}$ are LU equivalent.
\end{theorem}

\textit{Proof:---} Consider the local unitary invariants of
$\mathcal{H}^r=\{\ket{\varphi}^{\otimes r}:\ket{\varphi}\in \mathcal{H}\}$, where
these invariants are regarded as polynomials of $\op{\varphi}{\varphi}$ with
$\ket{\varphi}=\sum x_{j_1\cdots j_n}\ket{j_1\cdots j_n}$ and variables
$x_{j_1\cdots j_n}\in\mathbb{C}$. The set of local unitary invariants is the
intersection of rings $Q_i$, where $Q_i$ is the ring generated by the entries of
$tr_i\varphi^{\otimes r}=(tr_i\varphi)^{\otimes r}$.

We have the relation between the local unitary invariants of $\mathcal{H}^r$ and
the
LUIPs of the original system $\mathcal{H}$ as follows: Suppose
$f_1,f_2,\cdots,f_k$ of degree $2l_1,2l_2,\cdots,2l_k$ are LUIPs of the original
system. We have $f_j$ lies in the linear span of the entries of
$(tr_i\varphi)^{\otimes l_i}$. Then $\Pi_{j=1}^r f_j$ is a local unitary
invariants of $\mathcal{H}^r$ for the case $\sum l_k$ divisible by $r$. To see
this, we only need to observe that $\Pi_{j=1}^r f_j$ is an element of $Q_i$ since
$\Pi_{j=1}^r f_j$ can be generated by entries of $(tr_i\varphi)^{\otimes \sum
_{i=1}^k l_i}=[(tr_i\varphi)^{\otimes r}]^{\otimes \sum _{i=1}^k l_i/r}$, therefore, it can
be generated by entries of $(tr_i\varphi)^{\otimes r}$.

For any LUIP of the original system $\mathcal{H}$, $g$ with degree $2l$, we now
show that $g(\ket{\Psi})=g(\ket{\Phi})$.
To see this, we first choose a degree 2 LUIP $f_0$ be the square of 2-norm function, which satisfies that $f_0(\ket{\Psi})\equiv f_0(\ket{\Phi})$ according to the LU equivalence of $\ket{\Psi}^{\otimes r},\ket{\Phi}^{\otimes r}$. Then the following equation is valid with $i+l$ divisible by $r$,
\begin{equation*}
f_0^i(\ket{\Psi})g(\ket{\Psi})=f_0^i(\ket{\Phi})g(\ket{\Phi})
\end{equation*}
Therefore, $g(\ket{\Psi})=g(\ket{\Phi})$ is valid for any LUIP $g$.

By invoking the result that LUIPs are sufficient to separate any two distinct
orbits under local unitary, we can conclude that $\ket{\Psi},\ket{\Phi}$ are LU
equivalent.
\hfill $\blacksquare$

Similar statements are true for mixed states, and for quantum channels by
recalling the relation of LU equivalence between pure states, mixed states, and
unitary equivalence between quantum channels.

\section{SLOCC equivalence and SLIPs}
In the following, we provide an alternative algorithm to construct SLIPs.
It is direct to verify that we only need to compute the invariant polynomials of
group $\mathbb{SL}_i$, with $\mathbb{SL}_i=I_1\otimes \cdots\otimes
I_{i-1}\otimes\mathbb{SL}(d_i)\otimes I_{i+1}\cdots\otimes I_n$. We regard the
multipartite state as a bipartite pure state which is isomorphic to a matrix, says
$X$, and the action of the group is the left matrix multiplication, $i.e.$,
$X\rightarrow LX$ with $det(L)=1$. Fortunately, the invariant polynomials of such
map are fully characterized by the ring generated by the determinant of all square
matrix with columns catching from $X$, see \cite{KP96} for more details. Then, it
is direct to obtain the invariant ring of $\mathbb{SL}_i$, says $R_i$. After that,
we can present an algorithm to construct the ring of SLIPs for the multipartite
system $\mathcal{H}$, which is $\bigcap_{i=1}^n R_i$.

Observe that $R_i$ is generated by polnomials with degree $d_i$, the local dimension, then the degree of any element of $R_i$ is divisible by $d_i$.
One can confirm the following result which was obtained by Gour and Wallach in \cite{GW13}
\begin{observation}
Any SLIP has degree divisible by $lcm\{d_1,d_2,\cdots,d_n\}$, the least common multiple of ${d_1,d_2,\cdots,d_n}$.
\end{observation}

In the following, we study the equivalence between general mixed states under the action by SLOCC by employing local invariant polynomials.
Two quantum states $\rho$ and $\sigma$ of system $\mathcal{H}$ are called equivalent under one term SLOCC if there exists invertible $d_i\times d_i$ matrices $A_i$ such that
$$\rho=(A_1\otimes A_2\otimes \cdots)\sigma (A_1\otimes A_2\otimes \cdots)^{\dag}.$$
This definition captures the SLOCC equivalence between pure states and keeps the
tensor structure of the group, which enables us to characterize the local
invariants.
One can verify that
\begin{proposition}
$\rho$ and $\sigma$ are equivalent under one term SLOCC if and only if there is some $s\in \mathbb{SLU}$ such that $\ket{\Psi}$ is proportional to $s\ket{\Phi}$, where $\ket{\Psi},\ket{\Phi}\in \mathcal{H}\otimes\mathcal{H}_{n+1}$ are some purification of $\rho,\sigma$ with $d_{n+1}$ being the dimension of $\mathcal{H}_{n+1}$, and
$$
\mathbb{SLU}\equiv \mathbb{SL}(d_1)\otimes\mathbb{SL}(d_2)\otimes\cdots\otimes\mathbb{SL}(d_n)\otimes \mathbb{U}(d_{n+1}).
$$
\end{proposition}
Now we are ready to study the local invariant polynomials of $\mathbb{SLU}$: We define invariant polynomials of $\mathbb{SLU}$ as follows. A
function $f:\mathcal{H}\otimes\mathcal{H}_{n+1}\mapsto \mathbb{C}$ is an invariant polynomial of $\mathbb{SLU}$, if $f(\ket{\Psi})$ is
the homogenous polynomial on entries of $\Psi$, and
\begin{equation*}
f(s\ket{\Psi})=f(\ket{\Psi}), ~~\forall s\in \mathbb{SLU}~~\mathrm{and}~~\forall\ket{\Psi}\in \mathcal{H}\otimes\mathcal{H}_{n+1}.
\end{equation*}

Given $\ket{\varphi}=\sum x_{j_1\cdots j_{n+1}}\ket{j_1\cdots  j_{n+1}}$ with
variables $x_{j_1\cdots  j_{n+1}}\in\mathbb{C}$, one can notice that the invariant
ring, $T_{n+1}$, of the action of $\mathbb{U}(d_{n+1})$ is generated by
$tr_{n+1}{\varphi}$, some polynomials of entries of $\varphi$. For $i<n+1$, one
can obtain the invariant ring of $\mathbb{SL}_i$--$R_i$, as our previous argument
in the study of SLOCC equivalence between pure states. Note that one can not directly
compute the intersection of $R_i$ and $T_{n+1}$ since unlike $T_{n+1}$, $R_i$ is
generated by polynomials of $\ket{\varphi}$, not entries of $\varphi$. In order to
characterize the local invariant polynomials of $\mathbb{SLU}$, one should define
$T_i$ be the ring generated by elements of $R_i$ and $R^*_i$, with $R^*_i$
standing for the complex conjugate ring of $R_i$. Thus, $\bigcap_{i=1}^{n+1} T_i$
is the ring of local invariant polynomials of $\mathbb{SLU}$.

\section{Proof of Theorem 1}
In this section, we give a detailed proof of the upper bound on the degree of
generators of LUIPs (Theorem 1) and SLIPs respectively. Before doing so, we
recall the celebrated result of Derksen \cite{DER01} on the degree bounds in
invariant
theory.

Let $G$ be a linearly reductive
algebraic group over an algebraically closed field $K$ of characteristic
$0$, acting rationally
on an $s-$dimensional vector space $V$, specified as follows. $G$ is
given by
polynomials $h_1, \dots, h_\ell\in K[z_1, \dots, z_t]$ such that $G$ is the
zero set of these polynomials. The action of $G$ on $V$ is as follows:
there are polynomials $a_{i,j}$ for $i, j\leq s$, $a_{i, j}\in
K[z_1,\dots, z_t]$ such that
$g: G\to \mathbb{GL}(V)$ is given by $g\to (a_{i,j}(g))_{1\leq i,
j\leq s}$, where $\mathbb{GL}(V)$ is the general linear group of $V$, $i.e.$, the group of invertible matrices.

By fixing a basis of $V$, the polynomial functions over $V$ are identified as $R=K[x_1, \dots, x_s]$, and $G$ induces an action on $R$. The invariant ring of $G$ on $V$, denoted as $R^G$, consists of polynomials in $R$ invariant under $G$, $i.e.$,
\begin{equation*}
R^G=\{r:r(g\cdot v)=r(v),r\in R, \ \forall g\in G,v\in V\}.
\end{equation*}
It is known that $R^G$ is finitely generated.
The question here is to derive an
explicit degree bound for this finite generation. To obtain this degree bound,
another intermediate quantity is useful, and for this we recall the concept of
nullcone of $R^G$: it is defined as
the common zero set of all homogeneous polynomials in $R^G$ with
positive degree.

Let $\beta(V, G)$ be the minimal $k$ such that $R^G$ is generated by invariants of
degree less than $k$, and $\sigma(V, G)$ be the minimal $k$ such that the
invariants of degree less than $k$ defines the nullcone of $R^G$. Derksen shows
that \cite{DER01}
\begin{eqnarray*}
\sigma(V, G) \leq H^{t-d} A^{d}, \text{ and }\beta(V, G)\leq \max(2,\frac{3}{8}
s\cdot \sigma(V, G)^2),
\end{eqnarray*}
where $A=\max\{\deg(a_{i, j})\mid i, j\leq s\}$, $H=\max\deg(h_i)$, and
$d=\dim(G)$, the dimension of $G$ as an algebraic variety
\cite{DEF}.

To begin with, let us apply Derksen's bound to obtain an explicit upper bound for
the degree to generate the ring of SLIPs.
Let $\mathbb{S}=\mathbb{SL}(d_1,
\mathbb{C})\times \dots \times \mathbb{SL}(d_n, \mathbb{C})$ acts on
$\mathcal{H}$ in the natural way, where $\mathbb{SL}(d_i, \mathbb{C})$ is the
group of invertible $d_i\times d_i$ matrices with determinant 1. In this case
$R$ is a polynomial ring over $\mathbb{C}$ in $\prod_{i}d_i$ variables,
identified as the coordinate ring of $\mathcal{H}$. Our object is then the
invariant ring $R^{\mathbb{S}}$.

Note that
$\mathbb{S}$ is the zero locus of $\det(z_{i, j}^{(k)})_{i, j\in[d_k]}=1$,
$k=1, \dots, n$. In this setting, $t=\sum_{i}d_i^2$, $H=\max\{d_i\mid 1\leq i\leq
n\}$,
$\dim(\mathbb{S})=t-n$, and $A=n$.
Thus
$$
\sigma(\mathcal{H}, \mathbb{S})\leq \max(d_i)^n \cdot n^{\sum_{i}d_i^2-n}.
$$

As $s=\dim(R^{\mathbb{S}})\leq \prod_i d_i$, we get
$$
\beta(\mathcal{H}, \mathbb{S})\leq \frac{3}{8}\cdot (\prod_i d_i)\cdot
\max(d_i)^{2n}
\cdot
n^{\sum_i 2d_i^2-2n}.
$$
Therefore, the whole set of SLIPs can be generated by SLIPs with degree no more
than $\frac{3}{8}\cdot (\prod_i d_i)\cdot \max(d_i)^{2n} \cdot n^{\sum_i
2d_i^2-2n}$.

Now we prove Theorem 1.

\textit{Proof of Theorem 1:---}For a linear operation
$\rho$ in $\mathcal{H}$,
$g\in \mathbb{LU}\leq \mathbb{GL}(\mathcal{H})$ acts on
$\rho$ by sending
$\rho$ to $g \rho g^{\dagger}=g \rho g^{-1}$. Let $R$ be the polynomial ring
in $(\prod_{i}d_i)^2$ variables, identified as the coordinate ring of
$L(\mathcal{H}, \mathcal{H})$, and $R^{\mathbb{LU}}$ be LUIPs, $i.e.$, the
invariant ring of local unitary operations.

It is not feasible to apply Derksen's bound directly, as $\mathbb{U}$ cannot be viewed as
zero set of polynomials over algebraically closed field $\mathbb{C}$. This can be
fixed by considering the complexification. (For the notion of
complexification of compact groups, we refer the reader to \cite[Page
546]{Naimark}. For our purpose here, the concept of complexification associates
a compact connected semisimple Lie group with a semisimple connected complex
Lie group, s.t. their irreducible representations ``match.'')
In our case, the complexification of $\mathbb{U}$ yields
$\mathbb{G}=\mathbb{GL}(d_1,
\mathbb{C})\times \dots \times \mathbb{GL}(d_n, \mathbb{C})\leq \mathbb{GL}(\mathcal{H})$. Recall that we can view
$R$ as the space of representations of $\mathbb{LU}$ and $\mathbb{G}$,
and note that each invariant polynomial corresponds to the identity
representation.
Then by the correspondence between irreducible
representations of $\mathbb{LU}$ and $\mathbb{G}$, $R^\mathbb{LU}=R^\mathbb{G}$. Thus it is enough
to get a degree bound for the action of $\mathbb{G}$.

To get a degree bound for the action of $\mathbb{G}$ on $R$, we further notice
that for this particular action, the invariants of $\mathbb{G}$ and
$\mathbb{S}=\mathbb{SL}(d_1,
\mathbb{C})\times \dots \times \mathbb{SL}(d_n, \mathbb{C})$ coincide. This
allows us to apply Derksen's bounds to the group action of $\mathbb{S}$ as
follows.

Firstly $s=\dim(R^\mathbb{S})\leq \prod_{i}d_i^2$. To bound
$\sigma(\mathcal{H},\mathbb{S})$, we
observe that $t=\sum_i d_i^2$, $H=\max\{d_i \mid i=1, \dots, n\}$,
$d=\dim(\mathbb{S})=t-n$, and $A=\sum_{i}d_i$. Thus
\begin{eqnarray*}
 & &\sigma(L(\mathcal{H}, \mathcal{H}), \mathbb{G}) \\
 &=& \sigma(L(\mathcal{H}, \mathcal{H}), \mathbb{S})   \\
 & \leq & \max(d_i)^n\cdot (\sum_i d_i)^{\sum_id_i^2-n},  \\
&\Rightarrow & \beta(\L(\mathcal{H}, \mathcal{H}), \mathbb{G})\\
& =& \beta(L(\mathcal{H}, \mathcal{H}), \mathbb{S})  \\
 & \leq &  \frac{3}{8}(\prod_i d_i^2)\cdot \max(d_i)^{2n}\cdot (\sum_i d_i)^{2(\sum_i d_i^2-n)}.
\end{eqnarray*}\hfill $\blacksquare$

It is worth mentioning that the bounds present here is not optimal in general.

\section{LUIPs for two-qubit, three qubit system}
In this section, a set of local unitary invariant polynomials(LUIPs) for two-qubit system is obtained by using our method as an illustrating example.

For two-qubit system, let $\ket{\psi}_{AB}=\sum_{ij=0,1}x_{ij}\ket{ij}$, then
\begin{eqnarray*}
\rho_A=\left(
\begin{array}{cc}
 |x_{00}|^2+|x_{01}|^2 & x_{00}x_{10}^*+x_{01}x_{11}^*\\
 x_{00}^*x_{10}+x_{01}^*x_{11} & |x_{10}|^2+|x_{11}|^2\\
\end{array}
\right)~~~\rho_B=\left(
\begin{array}{cc}
 |x_{00}|^2+|x_{10}|^2 & x_{00}x_{01}^*+x_{10}x_{11}^*\\
 x_{00}^*x_{01}+x_{10}^*x_{11} &  |x_{01}|^2+|x_{11}|^2\\
\end{array}
\right).
\end{eqnarray*}
Let $S_A,S_B$ be sets of polynomials as,
\begin{eqnarray*}
S_A=\{|x_{00}|^2+|x_{01}|^2, x_{00}x_{10}^*+x_{01}x_{11}^*,x_{00}^*x_{10}+x_{01}^*x_{11}, |x_{10}|^2+|x_{11}|^2\},\\
S_B=\{|x_{00}|^2+|x_{10}|^2, x_{00}x_{01}^*+x_{10}x_{11}^*,x_{00}^*x_{01}+x_{10}^*x_{11}, |x_{01}|^2+|x_{11}|^2\}.
\end{eqnarray*}
According to the method of our main result, we can compute the LUIPs as following.

Degree 2 LUIP lies in the intersection of subspaces spanned by $S_A$ and $S_B$, one only need to deal with the following equation:
\begin{eqnarray*}
&&\alpha_0(|x_{00}|^2+|x_{01}|^2)+\alpha_1(x_{00}x_{10}^*+x_{01}x_{11}^*)+\alpha_2(x_{00}^*x_{10}+x_{01}^*x_{11})+\alpha_3(|x_{10}|^2+|x_{11}|^2)\\
&\equiv& \beta_0(|x_{00}|^2+|x_{10}|^2)+\beta_1(x_{00}x_{01}^*+x_{10}x_{11}^*)+\beta_2(x_{00}^*x_{01}+x_{10}^*x_{11})+\beta_3(|x_{01}|^2+|x_{11}|^2).
\end{eqnarray*}
By comparing the coefficients of this equation, we know that $\alpha_0=\alpha_3=\beta_0=\beta_3$ and $\alpha_1=\alpha_2=\beta_1=\beta_2=0$, which means that the only (up to a scalar) degree 2 LUIP is
$$\ip{\psi}{\psi}=\sum_{ij=0,1}|x_{ij}|^2.$$

The degree 4 LUIPs lies in the intersection of the subspaces spanned by $S_A^2$ and $S_B^2$, one only need to deal with the following equation:
\begin{eqnarray*}
&&\alpha_0(|x_{00}|^2+|x_{01}|^2)^2+\alpha_1(|x_{00}|^2+|x_{01}|^2)(x_{00}x_{10}^*+x_{01}x_{11}^*)+\alpha_2(|x_{00}|^2+|x_{01}|^2)(x_{00}^*x_{10}+x_{01}^*x_{11})\\
&+&\alpha_3(|x_{00}|^2+|x_{01}|^2)(|x_{10}|^2+|x_{11}|^2)+\alpha_4(x_{00}x_{10}^*+x_{01}x_{11}^*)^2+\alpha_5(x_{00}x_{10}^*+x_{01}x_{11}^*)(x_{00}^*x_{10}+x_{01}^*x_{11})\\
&+&\alpha_6(x_{00}x_{10}^*+x_{01}x_{11}^*)(|x_{10}|^2+|x_{11}|^2)+\alpha_7(x_{00}^*x_{10}+x_{01}^*x_{11})^2+\alpha_8(x_{00}^*x_{10}+x_{01}^*x_{11})(|x_{10}|^2+|x_{11}|^2)\\
&+&\alpha_9 (|x_{10}|^2+|x_{11}|^2)^2\\
&\equiv& \beta_0(|x_{00}|^2+|x_{10}|^2)^2+\beta_1(|x_{00}|^2+|x_{10}|^2)(x_{00}x_{01}^*+x_{10}x_{11}^*)+\beta_2(|x_{00}|^2+|x_{10}|^2)(x_{00}^*x_{01}+x_{10}^*x_{11})\\
&+&\beta_3(|x_{00}|^2+|x_{10}|^2)(|x_{01}|^2+|x_{11}|^2)+\beta_4(x_{00}x_{01}^*+x_{10}x_{11}^*)^2+\beta_5(x_{00}x_{01}^*+x_{10}x_{11}^*)(x_{00}^*x_{01}+x_{10}^*x_{11})\\
&+&\beta_6 (x_{00}x_{01}^*+x_{10}x_{11}^*)(|x_{01}|^2+|x_{11}|^2)+\beta_7(x_{00}^*x_{01}+x_{10}^*x_{11})^2+\beta_8(x_{00}^*x_{01}+x_{10}^*x_{11})(|x_{01}|^2+|x_{11}|^2)\\
&+&\beta_9(|x_{01}|^2+|x_{11}|^2)^2.
\end{eqnarray*}
By comparing the coefficients of this equation, we know that $\alpha_0=\alpha_9=\beta_0=\beta_9=\frac{\alpha_3+\alpha_5}{2}=\frac{\beta_3+\beta_5}{2}$ and $\alpha_1=\alpha_2=\alpha_4=\alpha_6=\alpha_7=\alpha_8=\beta_1=\beta_2=\beta_4=\beta_6=\beta_7=\beta_8=0$, which means that the degree 4 LUIPs are spanned by
\begin{eqnarray*}
&&(|x_{00}|^2+|x_{01}|^2+|x_{10}|^2+|x_{11}|^2)^2=tr^2(\rho_A)=tr^2(\rho_B),\\
&&(|x_{00}|^2+|x_{01}|^2)^2+(|x_{10}|^2+|x_{11}|^2)^2+2|x_{00}x_{10}^*+x_{01}x_{11}^*|^2\\
&=&(|x_{00}|^2+|x_{10}|^2)^2+(|x_{01}|^2+|x_{11}|^2)^2+2|x_{00}x_{01}^*+x_{10}x_{11}^*|^2=tr(\rho^2_A)=tr^2(\rho^2_B).
\end{eqnarray*}
It is well known that for two-qubit pure states, the degree 2 and 4 LUIPs can generate the whole ring of LUIPs.

\section{Conclusion}
In this paper, we give a characterization of multipartite entanglement by exploiting a systematic method to describe the ring of all LUIPs. More precisely, we then provide an algorithm to construct a set of generators of the ring of LUIPs. By employing our structure description of LUIPs, we are able to show that multipartite entanglement is additive in the sense that two multipartite states are LU equivalent if and only if $r$-copies of these two states are LU equivalent for some $r$. This idea gives an alternative way to study the multipartite entanglement in terms of equivalence classes of states under SLOCC, even for mixed states.

\vspace{0.1in} \noindent{\bf Acknowledgements.} We thank Prof. John Watrous, Prof. Debbie Leung and Dr. M. Grassl for their comments and Dr. S. Szalay's suggestion of references.
NY is supported by NSERC, NSERC DAS, CRC, CIFAR and DE180100156. JQ is supported by the Singapore Ministry of Education and the National Research Foundation and DE150100720. XS is supported by the National Natural Science Foundation of China Grant
61170062, 61222202.


\begin{thebibliography}{99}
\bibitem{DEF}Let $W$ be an algebraic set defined as the set of the common zeros of an ideal $I$ in a polynomial ring $R=K[x_1, \ldots, x_n]$ over a field $K$, and let $A=R/I$ be the algebra of the polynomials over $W$. Then the dimension of $W$ is:
    The maximal length d of the chains $V_0\subset V_1\subset \ldots \subset V_d$ of distinct nonempty subvarieties.
\end{thebibliography}
\end{document}